\documentclass[pra,twocolumn,showpacs,floatfix,a4paper]{revtex4}
\usepackage{graphicx}
\usepackage{hyperref}
\usepackage{epstopdf}
\graphicspath{{figure/}}

\newcommand{\ket}[1]{| #1\rangle}
\newcommand{\bra}[1]{\langle #1|}

\newcommand{\ketbra}[2]{\ket #1\bra #2}
\newcommand{\melc}[3]{\left\langle#1\left|#2\right|#3\right\rangle}

\begin{document}
\title{Implementation of a three-qubit quantum error correction code in a cavity-QED setup}
\author{Carlo Ottaviani}
\affiliation{Departament de F\'isica, Universitat Aut\`onoma de Barcelona, Bellaterra, E-08193, Spain}
\author{David Vitali}
\affiliation{School of Science and Technology, Physics Division, University of Camerino, Camerino, Italy}


\begin{abstract}
The correction of errors is of fundamental importance for the development of contemporary computing devices and of robust communication protocols. In this paper we propose a scheme for the implementation of the three-qubit quantum repetition code, exploiting the interaction of Rydberg atoms with the quantized mode of a microwave cavity field. Quantum information is encoded within two circular Rydberg states of the atoms and encoding and decoding process are realized within two separate microwave cavities. We show that errors due to phase noise fluctuations could be efficiently corrected using a state-of-the-art apparatus.
\end{abstract}

\maketitle

\section{Introduction}

Quantum computers offer the potential to solve certain classes of problems that appear to be practically unsolvable with classical computers. For example, they allow for efficient prime factorization \cite{shor}, and for the efficient simulation of quantum systems \cite{simulation}. However, quantum computers are particularly subject to the deleterious effects of noise and decoherence. In fact, the speed-up provided by quantum computers relies on the possibility to create and manipulate coherent superpositions of quantum states, which however are extremely sensitive to the coupling with external degrees of freedom. Therefore the protection from noise and errors is of fundamental importance for the realization of any quantum computer. At first sight, quantum error-correction seems to be precluded by the no cloning theorem \cite{noclon} which seems to rule out redundancy as usually employed in error correction. The discovery
of quantum error-correction codes (QECC) \cite{shor2,steane} that allow for fault tolerant quantum computing \cite{fault} has therefore made
the realization of practical quantum computers viable. The literature on the theory of QECC is vast and it now covers a wide range of approaches (see for examples the recent reviews of Refs.~\cite{munro,gottesman}). Instead experimental realizations have been limited to the field of liquid-state nuclear
magnetic resonance (NMR) \cite{qecc-nmr}, and to trapped ions \cite{chiaverini}. NMR experiments showed an increase in
state fidelity after performing the unitary operations of an error correction
protocol. However these first NMR demonstration have two drawbacks: i) NMR techniques cannot be scaled up with the number of qubits \cite{cory}; ii) the ancillary qubits
cannot be reset in these experiments, making therefore impossible in principle to repeat the protocol with the same qubits as many times as
required by a particular quantum algorithm. The trapped-ion implementation of a three-qubit QECC of Ref.~\cite{chiaverini} does not have instead these limitations, but it remains a unique example. Therefore studying the feasibility of the implementation of simple QECC protocols in alternative physical realizations of quantum computation is an important step for the development of the field.

In this paper we propose a scheme for the implementation of the three-qubit repetition QECC \cite{munro} on a cavity-QED setup. We show that one can show a significant increase of the state fidelity by implementing the scheme in state-of-the-art apparatus involving Rydberg atoms and microwave cavities \cite{haroche-nature}. Quantum information is encoded within circular Rydberg states and two cavities are employed to perform the encoding and decoding steps by means of standard cavity-QED techniques. In Sec.~II we review the main aspects of the three-qubit repetition code; in Sec.~III its specific implementation in the cavity QED setup is described. In Sec.~IV the main experimental features of the scheme are described, while in Sec.~V the results of the numerical simulations of the performance of the QECC protocol are presented and discussed. Sec.~VI is for concluding remarks.

\section{Basic Principles of Quantum Error Correction}

Let us assume that we have two distant stations, \emph{Alice} and \emph{Bob}, interested in sharing a message. This message can be modeled by a physical system that traveling through the communication channel is affected by unknown errors of various type \cite{ekert-mac,NielsenChuang}. The origin of these errors is the coupling of the system with the environment (\emph{decoherence}), and the ability in revealing and finding a way to correct them constitutes the central task of QECC.

Both classical and quantum error correction are based on the following three main stages: i) the \emph{encoding}, during which the original information is registered in a redundant way involving additional resources (the \emph{ancillas}); ii) the \emph{decoding}, in which the encoding process is reversed in order to distinguish which kind of error has taken place; iii) the \emph{correction} of the error for recovering the initial information. In the quantum case, the simplest way to encode information is to use a qubit and consider a generic state
\begin{equation}\label{eq:qubit}
\ket{\psi}_{s}=\alpha\ket{0}_{s}+\beta\ket{1}_{s},
\end{equation}\\
with $|\alpha|^{2}+|\beta|^{2}=1$. A simple and effective way of describing the degrading effect of the environment is by means of a collection of operators $\{\Gamma_{\mathcal{R}}^{(i)}\}$, acting on the qubit states $\ket{\psi}_{s}$, each associated with an environmental state $\ket{\psi}_{\mathcal{R}}^{(i)}$. For each $i$, the pair $\{\Gamma_{\mathcal{R}}^{(i)}\},\ket{\psi}_{\mathcal{R}}^{(i)}$ describes a type of error affecting the qubit. After preparing the message, the initial state of the qubit-environment system is the factorized state
\begin{equation}\label{eq:input-state-qecc}
\ket{\psi}_{in} = \ket{\psi}_{s}\otimes\ket{\psi}_{\mathcal{R}},
\end{equation}
which, after the disturbing action of the environment, becomes the entangled state
\begin{equation}\label{eq:output-state-qecc}
\ket{\psi}_{out} = \sum_{i}\ket{\psi}_{\mathcal{R}}^{(i)}\Gamma_{\mathcal{R}}^{(i)}\ket{\psi}_{s}.
\end{equation}
At the receiving station Bob does not have access to the environment variables and therefore the qubit state received by him is mixed and given by the trace over the environment
\begin{equation}\label{eq:bob-state}
\rho_{Bob} = Tr_{\mathcal{R}}(\ketbra{\psi_{out}}{\psi}).
\end{equation}\\
For a classical bit one can have only the flip error $|j\rangle \to |1-j\rangle $, $j=0,1$; instead the state of a qubit depends also upon the relative phase between $|0\rangle $ and $|1 \rangle $ and therefore one has two other independent errors: the phase error $|j\rangle \to (-1)^{j}|j\rangle $, and also its combination with the flip error $|j\rangle \to (-1)^{j}|1-j\rangle $. These three types of errors, associated with the three Pauli operators $\sigma_x$, $\sigma_y$ and $\sigma_z$, are all simultaneously present in a generic situation, and in order to correct for all of them the ``cheapest'' QECC requires at least four additional ancillary qubits \cite{NielsenChuang}. However, in many practical situations, an error type is much more relevant than the others and one can adopt the simpler \emph{three-qubit} QECC which is designed for correcting a single type of error \cite{NielsenChuang}. The three-qubit repetition code is the quantum extension of the repetition code \cite{NielsenChuang}, in which redundancy is obtained by circumventing the limitations imposed by the \emph{no-cloning theorem} by means of entanglement, i.e., by encoding the information of the initial qubit state of Eq.~(\ref{eq:qubit}) into the entangled state
\begin{equation}\label{eq:encoding-general}
\ket{\psi}_{S}=\alpha\ket{0}_{S}+\beta\ket{1}_{S}\rightarrow\ket{\psi}_{L}=\alpha\ket{0_{L}}+\beta\ket{1_{L}},
\end{equation}
where,
\begin{eqnarray}\label{eq:CQECC-enc}
\ket{0_{L}} &=& \ket{0}_{S}\otimes\ket{0}_{\mathcal{A}_{1}}\otimes\ket{0}_{\mathcal{A}_{2}}\\
\ket{1_{L}} &=& \ket{1}_{S}\otimes\ket{1}_{\mathcal{A}_{1}}\otimes\ket{1}_{\mathcal{A}_{2}},
\end{eqnarray}
with $\mathcal{A}_1$ and $\mathcal{A}_2$ denote the two ancillary qubits. The encoding process is performed by an entangling unitary operation $\mathcal{U}$ and after that the encoded state is affected by noise and errors introduced by the interaction with the environment. The decoding stage is then implemented by simply applying the inverse of the encoding operation, $\mathcal{U}^{\dagger}$, in order to determine which of the three qubits has been affected by the error. In fact, QECC theory assumes that the probability of having more than one error on a single qubit (between encoding and decoding) is negligible, i.e., that the coupling with the environment is weak. In such a limit the error can always be detected and the initial information perfectly recovered. The final correction stage can be performed in two different ways: i) ``automatically'', by a further unitary operation on the three-qubit system (a Toffoli gate \cite{NielsenChuang}) always yielding the original state of the qubit of interest; ii) by explicitly measuring the ancillas for detecting the error and eventually applying a feedback operation on the qubit for restoring the desired state. The first option is deterministic and therefore usually preferable, but the second option is also viable when highly efficient single qubit measurements are available and the implementation of three-qubit gates such as the Toffoli gate is too difficult (or too slow).
This latter scenario applies to the cavity-QED setup studied here, and therefore we shall consider from now on the ``measurement and feedback'' scheme for the implementation of the final correction stage.

\section{The cavity-QED implementation of the three-qubit QECC}

Many cavity-QED setups have now achieved the ability to perform entanglement and disentanglement operation with an high degree of accuracy \cite{haroche,kimble,rempe}. Here we shall focus in cavity-QED setup in the microwave regime \cite{haroche} which employs circular Rydberg atoms which represent a formidable tool for encoding quantum information due to their very long decay time (of the order of 30 ms).

\begin{figure}[t]
\begin{center}
\includegraphics[scale=0.35]{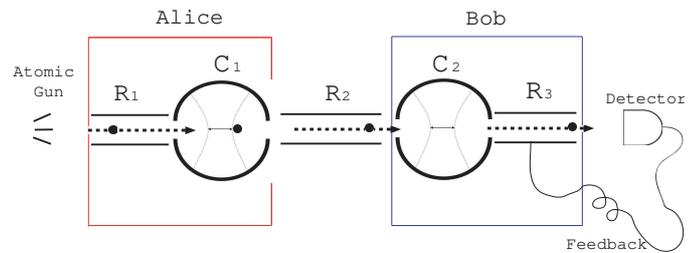}
\caption[Cavity QED scheme for Error Correction.]{Cavity-QED scheme of the proposed three-qubit Quantum Error Correction Code (QECC). The proposed set-up is analogous to the one currently developed at Ecole Normale Superi\'eure in Paris. It consists of three classical Ramsey zones ($R_{1,2,3}$) i.e., low-Q microwave cavities, and two high-Q microwave cavities ($\mathcal{C}_{1,2}$). Assuming that the length of the whole apparatus is $15$ cm, and that the atoms travel at $500$ m/s, the resulting total duration time of the protocol involving the four Rydberg atoms is $1.2$ ms.}\label{fig1}
\end{center}
\end{figure}

In order to implement the three-qubit QECC (see Fig.~\ref{fig1}) one needs two spatially separated high-Q microwave cavities, $\mathcal{C}_{1}$ and $\mathcal{C}_{2}$, in which the encoding and decoding process will be implemented, and four atoms $\mathcal{A}_{1},\mathcal{A}_{2},\mathcal{A}_{3},\mathcal{A}_{4}$, crossing both cavities and interacting sequentially with the each cavity mode.
The relevant atomic levels are three successive circular Rydberg levels with increasing energy, $\ket{i}, \ket{g}, \ket{e}$. The quantum information we want to protect is encoded in the first atom $\mathcal{A}_{1}$, while the second ($\mathcal{A}_{2}$) and the third atom ($\mathcal{A}_{3}$) are the ancillas, having the role of revealing the syndrome. Finally the fourth atom $\mathcal{A}_{4}$ is the atom on which the information, originally encoded in the the quantum state of $\mathcal{A}_{1}$, will be restored. Together with the two high-Q cavities, the scheme requires also three ``Ramsey'' zones ($\mathcal{R}_{i}$  for $i=1,\dots,3$ in Fig.~1) sandwiching the high-Q cavities, where classical microwave pulses can be applied for the manipulation of the atomic states. Finally one has a field-ionization atomic state detector \cite{haroche} which detects the error syndrome and activate the feedback correction loop. The cavities are assumed to be in the vacuum state at the beginning of the process; we also assume that the three circular Rydberg states can be prepared with high probability of success through the circularization process described in \cite{haroche}.
The QECC scheme proceeds through four steps which are now described.\\
\emph{First step: qubit state preparation. } The first atom $\mathcal{A}_{1}$ is prepared in level $\ket{e}$ and then enters cavity $\mathcal{C}_{1}$, which is initially in the vacuum state. The atom-cavity mode interaction is described in general by the following Jaynes-Cummings Hamiltonian ($\hbar=1$)
\begin{equation}\label{eq:H-cavity}
H = \Delta\hat{a}^{\dag}_{\mathcal{C}_{1}}\hat{a}_{\mathcal{C}_{1}} + \imath \frac{\Omega_{\mathcal{C}_{1}}}{2}(\hat{a}^{\dag}_{\mathcal{C}_{1}}\ketbra{g}{e} -\ketbra{e}{g}\hat{a}_{\mathcal{C}_{1}}),
\end{equation}\\
where $\Delta=\omega_{\mathcal{C}_{1}}-\omega_{ge}$ is the cavity detuning, which can be controlled in real time during the atomic passage via Stark shift tuning, i.e., by shifting in a controlled way the atomic levels through a static electric field applied in the cavity \cite{haroche}. In this case we consider perfect atom-cavity resonance, $\Delta =0$, so that the time evolved atom-cavity state is given by
\begin{equation}\label{eq:state_A1-C1}
\ket{e_{\mathcal{A}_{1}},0_{\mathcal{C}_{1}}}\rightarrow \cos{\frac{\Omega_{\mathcal{C}_{1}} t}{2}}\ket{e_{\mathcal{A}_{1}},0_{\mathcal{C}_{1}}} + \sin{\frac{\Omega_{\mathcal{C}_{1}} t}{2}}\ket{g_{\mathcal{A}_{1}},1_{\mathcal{C}_{1}}}.
\end{equation}\\
The atom-cavity interaction time $t$ can be also adjusted by using again Stark-shift tuning: in fact, the interaction can be stopped by shifting the $e \to g$ transition far-off resonance. In this way one can prepare an effective generic qubit state, encoded within the joint $\mathcal{A}_{1}$-$\mathcal{C}_{1}$ system,
\begin{equation}\label{eq:enc-state}
\ket{\psi}_{enc} = \alpha(t)\ket{e_{\mathcal{A}_{1}},0_{\mathcal{C}_{1}}} + \beta(t)\ket{g_{\mathcal{A}_{1}},1_{\mathcal{C}_{1}}},
\end{equation}\\
with $|\alpha(t)|^{2}+|\beta(t)|^{2}=1$.\\
\emph{Second step: the encoding stage}. We entangle the principal qubit with two atomic ancillas in order to obtain a state of the form given by Eq.~(\ref{eq:CQECC-enc}). The two atomic ancillas $\mathcal{A}_{2},\mathcal{A}_{3}$, are first prepared in the circular state $\ket{i}$. In the first Ramsey zone $R_{1}$ they undergo a $\pi/2$-pulse driving each ancilla to the superposition state
\begin{equation}\label{eq:ancillas-state-ini}
\ket{+_{\mathcal{A}_{2,3}}}=\frac{1}{\sqrt{2}}\left(\ket{i_{\mathcal{A}_{2,3}}}+\ket{g_{\mathcal{A}_{2,3}}}\right).
\end{equation}\\
The two ancillas are entangled with the prepared qubit state when crossing $\mathcal{C}_{1}$. The Stark shift field is set so that the $\ket{g}\rightarrow\ket{e}$ transition of both ancillas is resonant with the cavity mode. They both experience a $2\pi$ resonant cycle, so to realize a controlled phase accumulation on the atomic state $\ket{g}_{\mathcal{A}_{2},\mathcal{A}_{3}}$ resulting in $$\ket{1_{\mathcal{C}_{1}},g_{\mathcal{A}_{2,3}}}\rightarrow-\ket{1_{\mathcal{C}_{1}},g_{\mathcal{A}_{2,3}}}.$$\\ When both the ancillas have crossed $\mathcal{C}_{1}$, the final atoms-cavity state will be
\begin{equation}\label{eq:state-anc-C1}
\alpha(t)\ket{e_{\mathcal{A}_{1}},+_{\mathcal{A}_{2}},+_{\mathcal{A}_{3}},0_{C_{1}}} + \beta(t)\ket{g_{\mathcal{A}_{1}},-_{\mathcal{A}_{2}}-_{\mathcal{A}_{3}},1_{C_{1}}},
\end{equation}\\
where $\ket{\pm}_{A_{j}} = \frac{1}{\sqrt{2}}\big(\ket{i}_{A_{j}}\pm\ket{g}_{A_{j}}\big)$, with $j=2,3$.\\
\emph{Third step: the noisy channel}. The three encoded atoms travel from Alice to Bob through the noisy channel, modeled by a second Ramsey zone, $R_{2}$, where a random field is applied (for the details see section \ref{sec:channel}).\\
\emph{Fourth step: the decoding and the eventual correction}. We now describe the correction stage by considering in sequence the three possible situations: i) no error on the three atoms in $R_{2}$; ii) error on one of the ancillary atom ($\mathcal{A}_{2,3}$); iii) error on the encoded qubit ($\mathcal{A}_{1}$) (recall that the probability of two or more errors is assumed to be negligible).
\subsubsection{No error}
The decoding process takes place in the second cavity $\mathcal{C}_{2}$ which disentangles the three atoms. Let us first consider the case where there has been no error. Atom $\mathcal{A}_{1}$ interacts resonantly with a $\pi$-pulse with the cavity $\mathcal{C}_{2}$: the only part of the state 
that evolves is $\ket{e_{\mathcal{A}_{1}},0_{\mathcal{C}_{1}}}$, disentangling the first atom $\mathcal{A}_{1}$ from the rest, i.e.,
\begin{small}
\begin{equation}\label{eq:state-A1-C2}
\ket{\psi} = \left[\alpha(t)\ket{+_{\mathcal{A}_{2}},+_{\mathcal{A}_{3}},0_{\mathcal{C}_{1}},1_{\mathcal{C}_{2}}}+\beta(t)\ket{-_{\mathcal{A}_{2}},-_{\mathcal{A}_{3}},1_{\mathcal{C}_{1}},0_{\mathcal{C}_{2}}}\right]\ket{g_{\mathcal{A}_{1}}}.
\end{equation}
\end{small}\\
This transformation transfers the encoded qubit from the cavity-atom $\mathcal{C}_{1}-\mathcal{A}_{1}$ given by the relation (\ref{eq:state_A1-C1}) to the encoded qubit that involves the two cavities $\mathcal{C}_{1}-\mathcal{C}_{2}$. Then the two ancillas $\mathcal{A}_{2,3}$ pass through the second cavity, where they experience the same transformation they already experienced in $\mathcal{C}_{1}$, i.e., a $2\pi$-pulse resonant cycle. As a consequence, the two ancillas are decoupled from the encoded $\mathcal{C}_{1}-\mathcal{C}_{2}$ state as it must be for a decoding process, and the resulting state is an entangled state of the two cavities only,
\begin{equation}\label{eq:state-an-C2}
%
\ket{\psi} = \big[\alpha(t)\ket{0_{\mathcal{C}_{1}},1_{\mathcal{C}_{2}}} + \beta(t)\ket{1_{\mathcal{C}_{1}},0_{\mathcal{C}_{2}}}\big]\ket{-_{\mathcal{A}_{2}},-_{\mathcal{A}_{3}}}.
\end{equation}\\
Finally there is the final (eventual) correction stage: the ancillas are measured, and if an error is revealed the correction procedure is applied. In this step the fourth atom $\mathcal{A}_{4}$ starts to play its role: it has the function to reload the information now encoded in the $\mathcal{C}_{1}-\mathcal{C}_{2}$ entangled state and record the stored information in the atomic state. The need for this fourth atom is in fact evident, because the only information that can be efficiently read-out by the detector is that recorded in the atomic state.
Atom $\mathcal{A}_{4}$ is prepared in the circular state $\ket{g}$ and through Stark-shift tuning it is set far-off resonance from $\mathcal{C}_{1}$ so that it crosses it without interaction. Therefore it arrives in the same state $\ket{g}$ at $\mathcal{C}_{2}$, where it undergoes a resonant $\pi$-pulse interaction identical to that of atom $\mathcal{A}_{1}$. As a consequence, the state of the system becomes
\begin{equation}\label{eq:g4-C2}
\ket{\psi} = \big[-\alpha(t)\ket{0_{\mathcal{C}_{1}},e_{\mathcal{A}_{4}}} + \beta(t)\ket{1_{\mathcal{C}_{1}},g_{\mathcal{A}_{4}}}\big]\ket{-_{\mathcal{A}_{2}},-_{\mathcal{A}_{3}},0_{\mathcal{C}_{2}}}.
\end{equation}\\
The detection of the ancillary atoms provides the error syndrome: the two states $\ket{-}_{\mathcal{A}_{2}}\ket{-}_{\mathcal{A}_{3}}$ signals that the three qubits have not been affected by any error and therefore there is no correction to perform on the final qubit, atom $\mathcal{A}_{4}$.
Actually, the states of Eqs.~(\ref{eq:enc-state}) and (\ref{eq:g4-C2}) are not identical. Although the amplitude probabilities are exactly the same, they differ by a relative phase $\pi$. This is a consequence of the sequence of pulses realizing the entangling and disentangling operations, but there is a simple way to correct this problem, since it is sufficient to apply a classical $2\pi$-pulse resonant with the $i \to g$ transition in the $R_{3}$ zone. This will change the phase of the $g$ component of atom $\mathcal{A}_{4}$ only, permitting to obtain a perfect matching of the final and of the initially encoded state.
\subsubsection{Bit-flip error on the ancillas.}
A second possible option is that an error occurs on one of the ancillas $\mathcal{A}_{2,3}$. In this case the final state of Eq.~(\ref{eq:g4-C2}) will have one of the decoupled ancillary state flipped $\ket{-_{\mathcal{A}_{2}},-_{\mathcal{A}_{3}}}\rightarrow\ket{+_{\mathcal{A}_{2}},-_{\mathcal{A}_{3}}},\ket{-_{\mathcal{A}_{2}},+_{\mathcal{A}_{3}}}$. By detecting one of these two states of the ancillas, we can argue that the error has not involved the qubit we are sending, and therefore that, again, no correction is needed. As in the previous case, the only thing we have to do is to apply a classical pulse in $R_{3}$ to correct the $\pi $ relative phase.
\subsubsection{Bit-flip error on the encoded qubit.}
If instead an error occurs on $\mathcal{A}_{1}$ in $R_{2}$, the effect will be the exchange between $\ket{g}_{\mathcal{A}_{1}}\leftrightarrow\ket{e}_{\mathcal{A}_{1}}$. The state after the first atom has interacted with $\mathcal{C}_{2}$ now becomes
\begin{equation}\label{eq:g1-C2-bit-flip}
\ket{\psi} = \alpha(t)\ket{0_{\mathcal{C}_{1}},g_{\mathcal{A}_{1}}}\ket{+_{\mathcal{A}_{2}}, +_{\mathcal{A}_{3}}} + \beta(t)\ket{1_{\mathcal{C}_{1}},e_{\mathcal{A}_{1}}}\ket{-_{\mathcal{A}_{2}},-_{\mathcal{A}_{3}}},
\end{equation}\\
and after the passage of $\mathcal{A}_{4}$ we have
\begin{equation}\label{eq:g4-C2-bit-flip}
\ket{\psi} = \big[\alpha(t)\ket{0_{\mathcal{C}_{1}},g_{\mathcal{A}_{4}}} - \beta(t)\ket{1_{\mathcal{C}_{1}},e_{\mathcal{A}_{4}}}\big]\ket{+_{\mathcal{A}_{2}},+_{\mathcal{A}_{3}}}.
\end{equation}\\
In this case we have to apply the error correction, that consists of a feedback $\pi$ pulse in the first portion of the Ramsey zone $R_{3}$, flipping the state of $\mathcal{A}_{4}$ $\ket{g_{\mathcal{A}_{4}}}\leftrightarrow\ket{e_{\mathcal{A}_{4}}}$, followed by an off-resonant  $2\pi$ pulse, on the second portion of $R_{3}$, changing the phase of the $\ket{g_{\mathcal{A}_{4}}}$. The two pulses can be both performed within $R_{3}$ by using strong enough pulses. We finally obtain the following state (see eq.(\ref{eq:enc-state})) after the measurement of the two ancillas,
\begin{equation}\label{eq:psi_final}
\ket{\psi}_{final} = \alpha(t)\ket{0_{\mathcal{C}_{1}},e_{\mathcal{A}_{4}}} + \beta(t)\ket{1_{\mathcal{C}_{1}},g_{\mathcal{A}_{4}}},
\end{equation}\\
that is exactly the initial atom-cavity state provided $\mathcal{A}_{1}$ and  $\mathcal{A}_{4}$ are swapped.
The original encoded state is therefore restored and quantum information has been safely transferred from Alice to Bob.

\section{Experimental set-up, Rydberg states and Cavities}

The cavity scheme assumed in the previous Section fits well with the microwave-cavity setup at Ecole Normale Superi\'eure in Paris described, e.g., in \cite{haroche}. Let us now see in detail the properties of this setup and we show that the proposed QECC scheme can be implemented using a state-of-the-art apparatus, even when taking into account the experimental limitations due to spontaneous emission and the finite decay time of the cavities.
\subsection{The Circular Rydberg States}\label{sec:atoms}
Adopting atoms with long decay time and right velocities is of crucial importance for the realization of the protocol. Circular Rydberg states \cite{rydberg1} are excellent candidates because they correspond to large principal quantum number $n$ and maximum angular momentum $l=n-1$. The three level cascade structure can be found choosing atomic levels with principal quantum number $n=51,50,49$ for $\ket{e},\ket{g},\ket{i}$ respectively \cite{haroche}. The long radiative lifetime permits to have negligible effects on the atomic coherence from spontaneous emission, and the large dipole moment matrix elements, of the order of $1250$ a.u for the $\ket{e}\leftrightarrow\ket{g}$ transition, permits to have strong atom-field coupling. Circular Rydberg states can be prepared with a purity of $\ge98\%$ \cite{haroche-nature} and the velocity of the atoms in the atomic beam can be controlled with a precision of $\sim\pm 2$ m/s. The position of each atom inside the apparatus is known with a $\pm1$-mm precision.

\subsection{The Cavity}

The cavity \cite{haroche-nature,haroche} is an open Fabry-Perot resonator made with two spherical superconducting niobium mirrors facing each other at a distance $d=27.6$ mm, the diameter of the cavity is $D=50$ mm, and the radius of the mirrors is $R=40$ mm. The resonator is close to resonance with the $|e \rangle \leftrightarrow |g\rangle $ transition, with a maximum photon storage time of $T_{cav}\sim130$ms \cite{deleglise}, which corresponds to a record quality factor of $Q\sim3\times10^{8}$.
The vacuum state inside the cavity is obtained by cooling them down below $1$ K to avoid the presence of thermal photons. After cryogenic cooling, the mean photon number is still not negligible, $\sim0.7$, and the vacuum state is achieved with high fidelity by beginning every experiment with a flux of resonant atoms in the $\ket{g}$ state that, passing through the cavity, absorb the residual photons.

The time dependent coupling between the atoms and the cavity mode, $\Omega(t)$, is a Gaussian function depending on the atomic velocity $v$, that we set equal to $v=500$ m/s, and on the waist of the cavity mode, $w_{0} =6$ mm, and it is given by
\begin{equation}\label{eq:Gaussian-mode}
\Omega(t)=\frac{\Omega_{0}}{2}\mathcal{E}(t)=\frac{\Omega_{0}}{2}\exp\left[-\frac{v^2 t^{2}}{w_{0}^{2}}\right].
\end{equation}\\
As mentioned in previous Section, the interaction time can be controlled with high accuracy, thanks to Stark-shift tuning of the atoms injected inside the cavities \cite{haroche-nature,haroche}. A quick modification of the resonance conditions is possible by modulating the electric tension at the end of the two mirrors, resulting in a rapid change of electric field inside the cavity. This induces a quadratic Stark-shift of the atomic levels, that for Rydberg atoms is particularly strong \cite{nuss}. In this way all possible atom-cavity states of the form of Eq.~(\ref{eq:enc-state}) can be generated, by adjusting the accumulated Rabi angle $$\phi(t_{int})=\frac{\Omega_{0}}{2}\int_{-\infty}^{t_{int}}dt \mathcal{E}(t)\in[0,2\pi].$$

\section{Numerical Simulation}
The protocol has been simulated by choosing the parameters described in the previous Section and adopting the quantum trajectories (QT) approach \cite{qt-car} in order to solve for the time evolution of the atom-cavity system. We have considered two different initial states to encode, i.e., two different values of $\alpha(t)$ and $\beta(t)$. The QECC protocol is designed assuming perfect apparatus, i.e., a non-decaying atoms and cavities; the simulations includes these decay processes in order to verify to which extent the unavoidable imperfections and non-ideal features of the apparatus affect the efficiency of the algorithm.
\subsection{Simulation of the Noisy Channel}\label{sec:channel}
Let us now see in detail how the quadratic Stark shift effect can be used to engineer a noisy channel.
In the second Ramsey zone $R_{2}$ the atoms interact, for a controllable time, with a classical electromagnetic field. Applying a $\pi/2$ pulse to the $\mathcal{A}_{1}$ atom the encoded state at the exit of cavity $\mathcal{C}_{1}$ of Eq.~(\ref{eq:state-anc-C1}) becomes
\begin{equation}\label{eq:state-R2-1}
\ket{\psi} = \alpha(t)\ket{+_{\mathcal{A}_{1}},+_{\mathcal{A}_{2}},+_{\mathcal{A}_{3}},0_{C_{1}}}+
\beta(t)\ket{-_{\mathcal{A}_{1}},-_{\mathcal{A}_{2}}-_{\mathcal{A}_{3}},1_{\mathcal{C}_{1}}},
\end{equation}
so that the three qubits are all encoded in the $\ket{\pm}$ basis. A random electric field in $R_{2}$ generates, through the quadratic Stark shift effect, random phase shifts of the $|i\rangle$, $|g\rangle $, $|e\rangle $ states, which however become bit-flip errors in the $\ket{\pm}$ basis of the three qubits. This means that using the chosen encoding of Eq.~(\ref{eq:state-R2-1}) and random Stark-shifts, we engineer an effective channel in which each qubit is affected by the bit-flip error only.
After the application of the random electric field, we have to apply an inverse $\pi/2$ interaction on the $\mathcal{A}_{1}$ atom. This operation is needed in order to guarantee that, if there has been no error, the $\mathcal{A}_{1}$ state at the entrance of $\mathcal{C}_{2}$, when the qubits start to be processed by Bob, is identical to the state at the exit of $\mathcal{C}_{1}$. The random electric field in $R_2$ induces a shift of the energies of the three levels of interest, $\ket{i},\ket{g},\ket{e}$, due to quadratic Stark effect,
\begin{equation}\label{eq:stark-state-level}
\psi_{k}\rightarrow\psi_{k}\exp[-\imath T\Delta E_{k}/\hbar]\;\;k = i,g,e,
\end{equation}\\
where $T$ is the duration of the random electric field pulse in $R_2$. As a consequence, off-diagonal matrix elements with respect to atomic indices, \emph{i}, \emph{e}, \emph{g}, will acquire a random phase shift given by
\begin{equation}
\rho_{k,l}\to \rho_{k,l}\exp[-\imath T\left(\Delta E_k-\Delta E_l\right)/\hbar]\;\;k,l=i,g,e
\end{equation}\\
In the case of Rydberg levels, the energy shift $\Delta E_k$ due to the quadratic Stark shift caused by an electric field $\mathcal{E}$ is given by (in atomic units) \cite{nuss,arno}
\begin{eqnarray}\label{eq:random-stark-2}
\Delta E^{(2)} &=& -\frac{1}{8}\Big[7n^{2}-6(|m|^{2}+n_{1})^{2}+6n_{1}(|m|-1)\nonumber\\
&&+6n(|m|+1)-\frac{3}{2}|m|+8\Big]n^{4}|\mathcal{E}|^{2},
\end{eqnarray}
where $n_{1}$ is the parabolic quantum number, $n$ is the principal quantum number ($n=49,50,51$ for $\ket{i},\ket{g},\ket{e}$ respectively), and $m$ is the magnetic quantum number. For circular Rydberg states we have $n_1=0$ and $|m|=n-1$ so that
\begin{equation}\label{eq:energyshift}
\Delta E_n=-\frac{1}{8}\left[7n^2+\frac{21}{2}n+\frac{7}{2}\right]n^4 |\mathcal{E}|^2 \equiv \alpha_n |E|^2.
\end{equation}
As a consequence, the phase shift of an off-diagonal matrix element due to the application of the random electric field can be written as
\begin{eqnarray}
\rho_{k,l}&\to& \rho_{k,l}\exp\left[-\imath T |\mathcal{E}|^2 \left(\alpha_k-\alpha_l\right)/\hbar\right]\nonumber\\
&=&\rho_{k,l}\exp\left[-i\phi \left(\alpha_k-\alpha_l\right)\right], \label{eq:phaseshift}
\end{eqnarray}\\
where $\phi$ is a random phase proportional to the intensity of the Stark field and which we shall assume as uniformly distributed over an interval $[0,\phi_{max}]$. Therefore one has random state-dependent Stark shifts determined by Eqs.~(\ref{eq:energyshift}) and (\ref{eq:phaseshift}).

\subsection{Discussion of the results}

The state of the whole systems evolves in a Hilbert state of dimension $n=324$ (four three-level atoms and two cavities with one photon at most). The adoption of the QT approach permits to manipulate the evolution of a wave function and not of a density matrix, as it happens when solving master equations, which instead would have implied working in a much larger space of dimension $324^2$.
The density matrix of the whole system is obtained by averaging over the trajectories, and in our case we have
simulated the proposed QECC protocol and performed the average over 1000 trajectories.
During each trajectory a flat-distributed random phase $\phi$ chosen in the interval $[0,\phi_{max}]$ on the three atoms is induced. The performance of the QECC protocol is analysed by comparing the fidelity \cite{NielsenChuang}
\begin{equation}\label{eq:fid-err-est}
\mathcal{F} = \sqrt{\melc{\psi_{enc}}{\rho_{final}}{\psi_{enc}}},
\end{equation}
in the two cases, i.e., when QECC is applied and when we do not complete the final correction step, i.e. we do not perform the feedback action in $R_{3}$ on the fourth atom.
We always find a clearly visible difference between the two fidelities, showing the validity of the protocol even in the presence of a non-ideal apparatus. Both the corrected and the uncorrected fidelity are oscillating function of the error strength, which we measure in terms of the maximum possible random phase shift, $\phi_{max}$, and both tend to an asymptotic value for large values of $\phi_{max}$.

\begin{figure}[t]
\begin{center}
\includegraphics[scale=0.35]{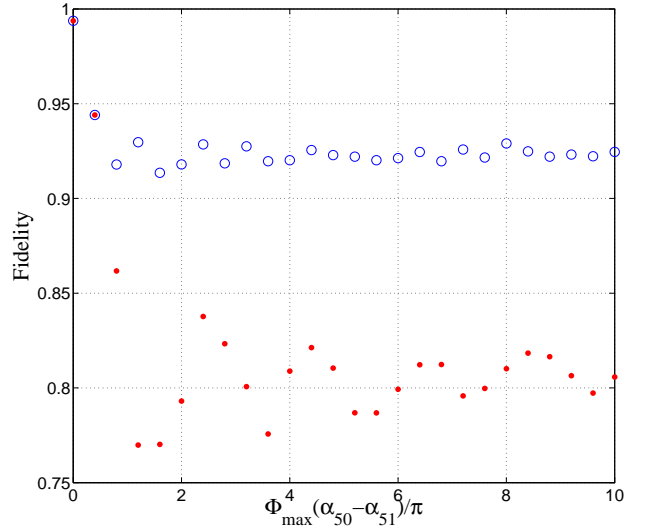}
\caption[Fidelity of the QECC scheme based on the CQED.]{(Color online) Fidelity of the proposed QECC scheme. Parameters are those of Ref.~\protect\cite{haroche-nature}, corresponding to a cavity decay time $T_{cav}=100$ ms. The initial state to protect has $\alpha=\sqrt{0.7}$ and $\beta=\sqrt{0.3}$. Blue circles correspond to the QECC scheme, while red dots correspond to the case without correction.}\label{fig2}
\end{center}
\end{figure}

Fig.~\ref{fig2} refers to an initial state to protect equal to $\sqrt{0.7}\ket{e_{\mathcal{A}_{1}},0_{\mathcal{C}_{1}}} + \sqrt{0.3}\ket{g_{\mathcal{A}_{1}},1_{\mathcal{C}_{1}}}$, and to a parameter choice corresponding to those of the experiment of Ref.~\cite{haroche-nature}, which means in particular a cavity decay time $T_{cav}=100$ ms. Without QECC, the fidelity $\mathcal{F}$ tends to an asymptotic value for large phase shifts around $\mathcal{F} \sim 0.8$, while it tends to $\mathcal{F} \sim 0.92$ in the presence of QECC. In Fig.~3 we study how the performance of the QECC protocol depends upon the chosen initial state, by comparing the fidelity with and without QECC for the states with $\alpha = \sqrt{0.7}$ or $\alpha = \sqrt{0.6}$. We see that this dependence is extremely weak, especially in the presence of error correction. Finally in Fig.~4 we study the dependence of the protocol performance upon the imperfection of the apparatus, and upon the cavity decay time in particular, by comparing the cases with $T_{cav}=100$ ms and $T_{cav}=1$ ms. As expected, the performance worsens for shorter cavity decay time, but again the dependence is very weak and one has only a small decrease of the fidelity for microwave cavities with a lifetime $100$ times shorter. These results show the robustness of the proposed QECC protocol, which provide a significant state protection even in the presence of not negligible loss processes.

The effect of cavity and atomic decay is well visible by looking at the data points corresponding to no error, $\phi_{max}=0$: the fidelity both with and without QECC is not equal to one, due to the fact that the evolution is not unitary. Cavity decay and atomic spontaneous emission, even though small, are not zero, determining a nonzero error probability. However, even in the case of $T_{cav}=1$ ms, such error probability is small because decay processes are still much slower than typical interaction times, which in our case can be taken of the order of $20\mu$s.

\begin{figure}
\begin{center}
\includegraphics[scale=0.35]{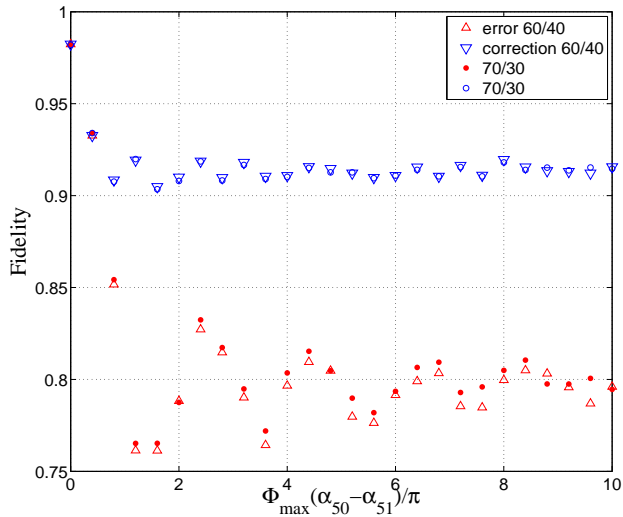}
\caption[Fidelity of the QECC scheme based on the CQED.]{(Color online) Comparison between two different initial states: i) $\alpha=\sqrt{60}$, $\beta=\sqrt{40}$, (red triangles, no correction, and blue triangles, with QECC); ii) $\alpha=\sqrt{70}$, $\beta=\sqrt{30}$ (red dots, no correction, and blue circles, with QECC). The cavity decay time is $T_{cav}=1$ ms, while the other parameters are the same as those of Fig.~2.}\label{fig3}
\end{center}
\end{figure}
\begin{figure}
\begin{center}
\includegraphics[scale=0.35]{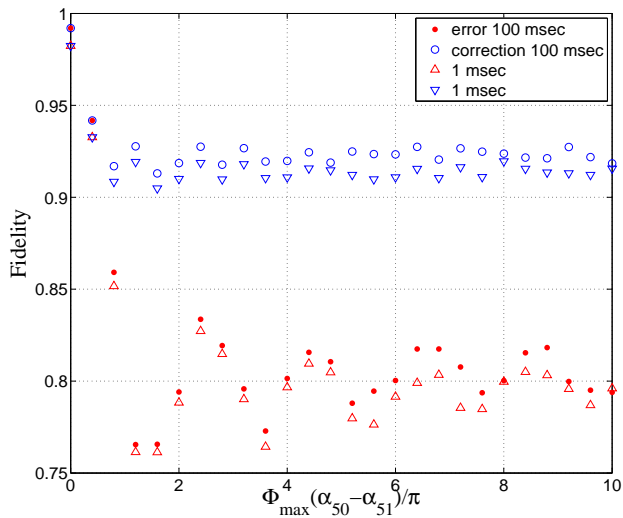}
\caption[Comparison for two different decay rate of the cavities]{(Color online) Comparison for two different cavity decay times $T_{cav}=100$ ms (blue circles for the QECC case and red dots with no QECC), and $T_{cav}=1$ ms (blue triangles for the QECC case and red triangles with no QECC). In both cases the initial state is with $\alpha=\sqrt{60}$ and $\beta=\sqrt{40}$. The other parameters are the same as in Fig.~\protect\ref{fig2}}\label{fig4}
\end{center}
\end{figure}

\begin{figure}
\begin{center}
\includegraphics[scale=0.6]{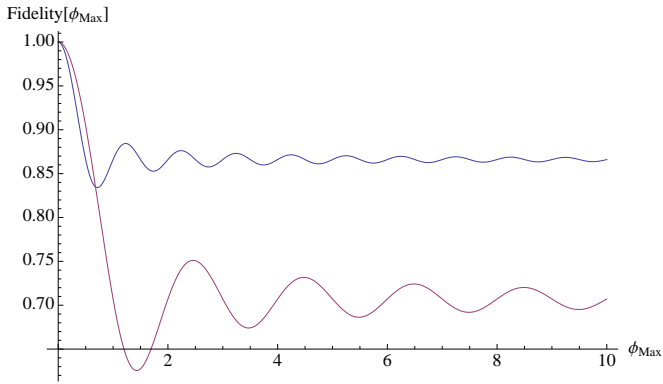}
\caption[]{Fidelities associated with the simple model of Eqs.~(\ref{eq:nofeed}) and (\ref{eq:feed}) with (upper curve, blue) and without correction (lower curve, red).}\label{fig6}
\end{center}
\end{figure}

These results can be qualitatively explained by adopting a simple model for the correction protocol. In practice, we want to transfer a given state from Alice to Bob by crossing a noisy region. Without loss of generality we can always choose the basis so that the state we want to recover at Bob site is called $|e\rangle$. The Stark random phase $\phi$ together with the two $\pi/2$ pulses (the operations described in Sec.~\ref{sec:channel}) in $R_2$ is equivalent to a rotation by a random angle $\phi$ of the transferred state. The state at Bob site can be therefore written as
$$
|\psi\rangle= \cos\frac{\phi}{2}|e\rangle + \sin\frac{\phi}{2}|g\rangle.
$$
The measurement of the ancillas is practically equivalent to a measurement in the e-g basis, in order to check if the desired state $e$ has arrived at destination or not. Therefore the probability of success of the transport, without any correction, is simply $\cos^2\phi/2$ and the corresponding fidelity, is simply given by
\begin{equation}
F_{nofb}=\left|\cos\frac{\phi}{2}\right|.
\end{equation}
In this simple model, the correction after the measurement is described by the application of the spin flip operator $\sigma_x = |e\rangle \langle g |$ whenever one detects $g$, which occurs with probability $P_{err}=\sin^2\phi/2$, and no correction in the other cases. The resulting final state is a mixed state, given by
\begin{equation}
\rho_{fin}= P_{err} \sigma_x |\psi\rangle \langle \psi | \sigma_x + (1-P_{err})|\psi\rangle \langle \psi |.
\end{equation}
The success probability in the case of feedback is $\langle e |\rho_{fin} |e\rangle$ and taking the square root, one gets
\begin{equation}
F_{fb}=\sqrt{\cos^4\frac{\phi}{2}+\sin^4\frac{\phi}{2}},
\end{equation}
The phase $\phi$ is random, and one has to average these results over a flat distribution between zero and $\phi_{max}$. Just to simplify the analytical calculation, instead of averaging the fidelity (i.e., the average of the square root of the success probability), the result can be approximated by the square root of the average of the success probability. One gets
\begin{eqnarray}
&&F_{nofb}^{ave} \simeq \left[\frac{1}{\phi_{max}}\int_0^{\phi_{max}} d\phi \cos^2\frac{\phi}{2}\right]^{1/2} \nonumber \\
&& =\left[\frac{1}{2}+\frac{\sin\phi_{max}}{2\phi_{max}}\right]^{1/2},\label{eq:nofeed} \\
&&F_{fb}^{ave} \simeq \left[\frac{1}{\phi_{max}}\int_0^{\phi_{max}} d\phi \left(\cos^4\frac{\phi}{2}+\sin^4\frac{\phi}{2}\right)
\right]^{1/2} \nonumber \\
&& =\left[\frac{3}{4}+\frac{\sin 2\phi_{max}}{8\phi_{max}}\right]^{1/2}.\label{eq:feed}
\end{eqnarray}
Fig.~\ref{fig6} shows the resulting fidelity as a function of the error strength. The two curves manifest the same qualitative behavior of the numerical results shown in Figs.~2-4. The simplified expressions of Eqs.~(\ref{eq:nofeed})-(\ref{eq:feed}) however underestimates both fidelities and also the performance of the QECC scheme, predicting lower asymptotic values of the fidelities. This is related to the fact that we have overestimated the effect of errors, because we have assumed that the errors mainly affects the qubit of interest $\mathcal{A}_1$ and neglected errors occurring on the ancillas.

\section{Conclusions}

We have proposed a scheme for the implementation of the three-qubit QECC using a cavity QED setup. In particular we have considered a state-of-the-art apparatus in which the quantum information to protect is encoded in the state of circular Rydberg atoms crossing two high-quality microwave cavities. The encoding and decoding steps of the three-qubit QECC are performed within the cavities exploiting a resonant atom-cavity interaction and the possibility to manipulate this interaction by Stark shifting the atomic levels. The error syndrome and the eventual correction is performed by explicitly detecting the atomic state and via a controlled $\pi$-pulse operation. By considering the same parameter regime of the recent experiment of Ref.~\cite{haroche-nature} we have shown that QECC significantly preserves the prepared atomic state against the noise due to fluctuating electric fields which randomly shift the atomic energy levels via quadratic Stark effect. Our analysis has taken into account all the major experimental limitations, i.e., cavity decay and spontaneous emission, and we have seen that the performance of the QECC is not too much affected by them, as long as the decay times are much longer than the typical atom-cavity interaction times. Instead, we did not take into account the non-unit efficiency of atomic detectors. In fact, as long as the detection efficiency does not depend on the atomic state, the fact the probability of missing an atom is nonzero does not affect the efficiency of the protocol, but only decreases the rate of significative events in performing the experiment.

The present scheme can be extended for implementing more involved quantum error-correction codes, e.g. the five- or seven-qubit code, paying only the price of a slightly more involved sequence of operations. In fact, the main limitation for scaling-up the scheme to more qubits is represented by the spontaneous emission of the circular Rydberg states, which is of the order of 30 ms and limits the number of atoms that can be prepared and manipulated. Cavity decay time is less important because the cavity modes are used only for much shorter time for carrying out the operations. Assuming an atomic velocity around $500$ m/s (as used in our simulations) the time duration of the while protocol is around $1.2$ ms and therefore there is enough time for scaling up to at least ten qubits. In order to scale-up to a realistic and useful quantum processor the present cavity-QED setup is less suitable. In fact, one should avoid spontaneous emission by encoding quantum information in hyperfine-split ground-state sublevels and by employing optical Raman transitions between these levels. Replacing microwave cavities and transitions with optical ones has also the advantage of avoiding cryogenic setups. For example, scalable cavity-QED configurations of this kind could be provided by coupled cavity arrays \cite{cca}.

\section{Acknowledgements}

We thank Michel Brune and Jean Michel Raimond for illuminating discussions and suggestions. This work has been partially supported by the EC FP-6 IP project SCALA and by the EC FP-7 STREP project HIP. Carlo Ottaviani is also supported by the Juan-de-la-Cierva Grant of the ``Spanish Ministerio de Educaci\'on y Ciencia''.

\end{document}